\documentclass[%
 reprint,
 amsmath,amssymb,
 aps,
]{revtex4-2}

\usepackage[utf8x]{inputenc}
\usepackage[T1]{fontenc}
\usepackage{amsmath}
\usepackage{mathtools}
\usepackage{graphicx}
\usepackage{comment}
\usepackage{placeins}
\usepackage{bm}
\usepackage{color}

\usepackage{amssymb}
\usepackage{booktabs}
\usepackage{braket}
\usepackage{float}
\usepackage{glossaries}
\usepackage{tabularx}
\usepackage{units}
\usepackage[version=4]{mhchem}
\usepackage[dvipsnames]{xcolor}
\usepackage[colorlinks=true,
            linkcolor=NavyBlue,
            citecolor=NavyBlue,
            filecolor=NavyBlue,
            urlcolor=NavyBlue
            ]{hyperref}

\graphicspath{{./figures/}}

\renewcommand{\vec}[1]{\ensuremath\boldsymbol{#1}}

\newcommand{\corr}{\textcolor{black}}
\newcommand{\corrmb}{\textcolor{black}}
\setacronymstyle{long-short}
\newacronym{fdtd}{FDTD}{finite-difference time-domain}
\newacronym{ed}{ED}{electric dipole}
\newacronym{eq}{EQ}{electric quadrupole}
\newacronym{eo}{EO}{electric octupole}
\newacronym{md}{MD}{magnetic dipole}
\newacronym{mq}{MQ}{magnetic quadrupole}
\newacronym{mo}{MO}{magnetic octupole}
\newacronym{hbn}{hBN}{hexagonal boron nitride}
\newacronym{hns}{HNS}{hyperbolic nanosphere}
\newacronym{vdw}{vdW}{van der Waals}
\newacronym{tmd}{TMD}{transition metal dichalcogenide}
\newacronym{si}{SI}{Supporting Information}
\newacronym{qs}{QS}{quasistatic}
\newacronym{rms}{RMS}{root mean square}
\newacronym{pvd}{PVD}{physical vapor deposition}
\newacronym{ald}{ALD}{atomic layer deposition}

\newcommand{\warsaw}{
    Faculty of Physics,
    University of Warsaw,
    Pasteura 5,
    02-093 Warsaw, Poland
}
\begin{document}

\title{Polarization-dependent mode coupling in hyperbolic nanospheres}
% Optical modes  of spherical hyperbolic nanoresonators
% Hyperbolic dispersion begets coupling of electric and magnetic modes in optical nanoresonators
% Polarization-dependent mode coupling in hyperbolic nanospheres

\author{Krzysztof M. Czajkowski}
\thanks{These authors contributed equally.}
\affiliation{\warsaw}

\author{Maria Bancerek}
\thanks{These authors contributed equally.}
\affiliation{\warsaw}

\author{Alexander Korneluk}
\affiliation{\warsaw}

\author{Dominika \'Switlik}
\affiliation{\warsaw}

\author{Tomasz J.\ Antosiewicz}
\affiliation{\warsaw}
\email{tomasz.antosiewicz@fuw.edu.pl}
%\phone{+48 22 55 32 006}

\date{\today}

%\begin{tocentry}
%\end{tocentry}

\begin{abstract}
  Hyperbolic materials offer a much wider freedom in designing optical properties of nanostructures than ones with isotropic and elliptical dispersion, both metallic or dielectric.  Here, we present a detailed theoretical and numerical study of the unique optical properties of spherical nanoantennas composed of such materials. Hyperbolic nanospheres exhibit a rich modal structure that, depending on the polarization and direction of incident light, can exhibit either a full plasmonic-like response with multiple electric resonances, a single, dominant electric dipole or one with  mixed magnetic and electric modes with an atypical reversed modal order. We derive resonance conditions for observing these resonances in the dipolar approximation and offer insight into how the modal response evolves with the size, material composition, and illumination. Specifically, the origin of the magnetic dipole mode lies in the hyperbolic dispersion and its existence is determined by two diagonal permittivity components of different sign. Our analysis shows that the origin of this unusual behavior stems from complex coupling between electric and magnetic multipoles, which leads to very strongly scattering or absorbing modes. These observations assert that hyperbolic nanoantennas offer a promising route towards novel light-matter interaction regimes.
\end{abstract}

\maketitle

\section{Introduction}

\noindent
The optical response of a particular system to external illumination is governed not only by the oscillator strength of the systems' electrons \cite{Yang2015}, but also by their spatial distribution. Indeed, structure alongside composition are the two knobs which determine a desired response or functionality.
At the single-particle level, the initial choice is the use of metals or dielectrics. The surface plasmon-polariton, an inherent material resonance, uses very efficiently the collective response of many conduction electrons to shape the electromagnetic field at the nanoscale, enabling, for example, plasmon-enhanced photochemistry \cite{Iandolo2013-PCCP}, subwavelength confinement of light \cite{BaumbergNature2016}, single-molecule biosensors \cite{Acimovic2018}, and antennas for detection of volatile chemical species \cite{2019_NatMater_18_489_ferry}. 
An equally rich optical response is possible with dielectric nanostructures, however, theirs is fundamentally a geometrical resonance. They simultaneously support both electric and magnetic resonances and can be chosen to have small or negligible dissipation \corr{\cite{Garcia-Etxarri2011}. These were subsequently observed in spherical nanoparticles \cite{Kuznetsov2012, Evlyukhin2012}, whose response is characterized by a high degree of symmetry and allows to capture the fundamental properties of scatterers made of a particular type of material.}
The presence of both types of multipoles allows for significant tunability of the optical response already in simple structures \cite{Evlyukhin2011, Staude2013-ACSNano}.
Some of the more spectacular effects include unidirectional scattering from nanodisks via generalized Kerker effects \cite{Liu2018}, leading to high-efficiency Huygens metasurfaces \cite{2015_AdvOptMater_3_813_decker} or magnetic mirrors \cite{2014_ApplPhysLett_104_171102_moitra}. 

The above-mentioned examples of physical effects and/or applications are obtained by tuning the size and shape of plasmonic or dielectric nanostructures \cite{Evlyukhin2011}, all the while utilizing isotropic materials \cite{Giannini2011, Evlyukhin2020}. 
In such structures, the optical response can be made strongly dependent on the direction of incident light and/or its polarization \cite{Kats2012}, however, even more freedom in designing a desired functionality can be obtained by using anisotropic materials. At the very basic level an illumination/polarization-dependent optical response can be achieved not by tuning the shape of the particle \cite{2014_NatureComm_5_1135_ross, Persechini_2014}, 
but rather the birefringence of the constituent material \cite{2016_OptLett_41_3563_liu, Ermolaev2021}.

A particularly interesting class of anisotropic materials are hyperbolic ones, whose permittivity (or in principle permeability) tensor is not only diagonal, but one of the principal components is opposite in sign to the other two, leading to \corrmb{unbounded, hyperbolic isofrequency surfaces and supporting high-\textit{k} modes} \cite{2013_NatPhoton_7_958_poddubny}. 
Hyperbolic metamaterials are presently an active research field, with investigated topics encompassing spontaneous emission \cite{LopezMorales2021}, waveguides \cite{Roth2017}, subdiffraction focusing and imaging \cite{Dai2015}, and others. 
This type of dispersion can be found in naturally existing \gls{vdw} materials \cite{Hu&Shen2020} such as \gls{hbn} \cite{2014_NatCommun_5_5221_caldwel}, \ce{WTe2} \cite{2020_NatureCommun_11_4592_wang} and current research is directed towards searching for new ones \cite{Choe2021}. However, a viable alternative to obtaining hyperbolic dispersion is to use artificial metal-dielectric structures such as continuous film \cite{Liu2007} or fishnet-type \cite{Kruk2016} multilayers or nanorod arrays \cite{Roth2017}.

\begin{figure*}
  \includegraphics[width=0.9\textwidth]{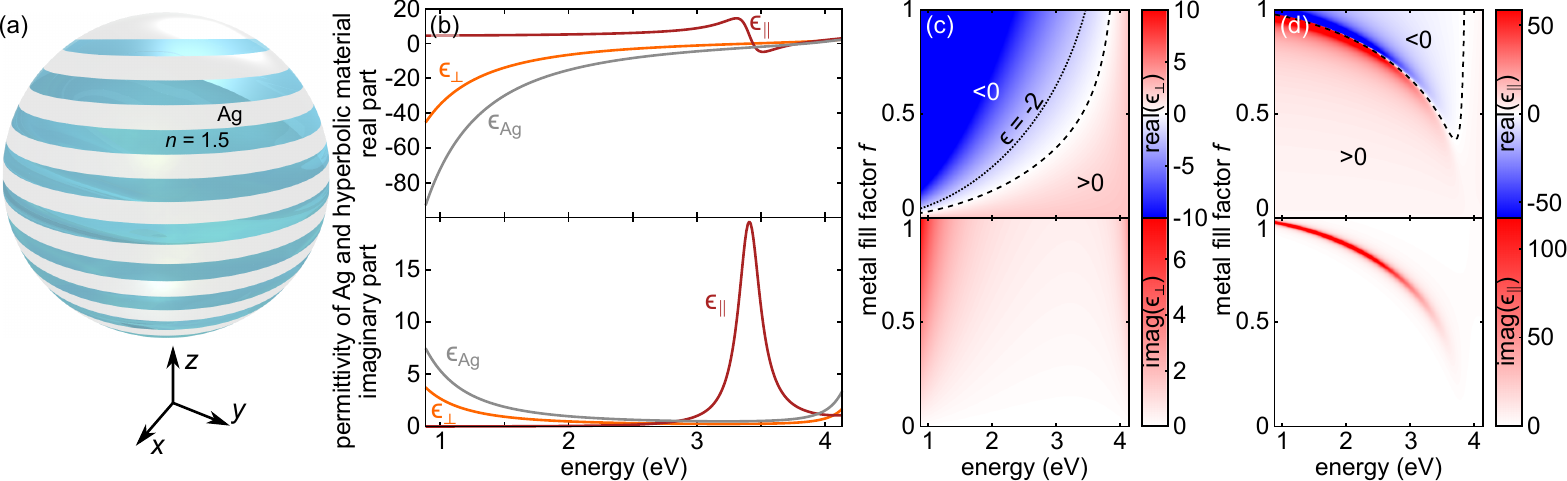}
  \caption{(a) Scheme of a uniaxial spherical nanoparticle which is composed of alternating dielectric ($n_d=1.5$) and silver layers. In the chosen coordinate system we then define the parallel $\epsilon_{xx}=\epsilon_{yy}\equiv\epsilon_{\perp}$ and perpendicular $\epsilon_{zz}\equiv\epsilon_{\parallel}$ permittivities which for a metal fill factor $f_{m}=0.5$ are plotted in panel (b). The permittivity of silver, based on Palik,\cite{palik_vol1} is plotted for reference. For energies below ca. \unit[3.5]{eV} the nanosphere is characterized by a hyperbolic dispersion relation. (c,d) Real and imaginary parts of permittivity $\epsilon_\parallel$ and $\epsilon_\perp$ used throughout this work as function of $f_m$. The dashed lines mark 0 and split $\epsilon(\hbar\omega,f_m)$ into different dispersion regimes.}
  \label{fig:scheme}
\end{figure*}

The novel functionalities of large-scale hyperbolic materials \cite{2013_NatPhoton_7_958_poddubny} can be extended or adapted to nanoparticle-based optical cavities. One of these properties is confinement of optical modes, which was initially realized in stacked Ag/Ge multilayers \cite{Yang&Yao2012}, although due to large losses the quality factor of the resonances was quite low. This can be alleviated by employing low-loss materials, such as nanostructures composed of \gls{hbn}, which features two restrahlen bands with hyperbolic dispersion of different type in the mid-IR. While the low-losses of \gls{hbn} are very beneficial and transient tuning of permittivity is possible \cite{Sternbash2021}, use of natural hyperbolic \gls{vdw} materials restricts the parameter space of material properties. \corr{However, recent advances in stacking \gls{tmd} heterostructures \cite{Wang2021}  hint at the possibility of overcoming such limitations. Alternatively, at-fabrication tunability of the birefringence of hyperbolic materials is enabled by use of engineered metal/dielectric multilayers via various lithographic techniques that utilize \gls{pvd} methods to alternately deposit both materials \cite{Maccaferri2019} or etch nanostructures in prepared multilayers \cite{Wang2003}}. Indeed, optical type II hyperbolic nanoantennas have been recently shown to exhibit simultaneously radiative and non-radiative modes \cite{Maccaferri2019} as well as offer enhanced photoluminescence \cite{Indukuri2020} and nonlinear emission \cite{Maccaferri2021}.

The above-discussed interest in hyperbolic nanoantennas and potential use in optical devices makes it necessary to understand, at the fundamental level, the behavior of optical modes of these structures. While discussion on basic properties of anisotropic resonators can be found in the literature \cite{Hofer2020, Kossowski2021} and is relevant to the topic at hand, hyperbolic nanoparticles need separate treatment.
Here, we contribute to this topic by discussing basic electromagnetic properties of \glspl{hns}. \corr{The motivation behind studying a spherical object is the unambiguity of the hyperbolic origin of the optical properties, which are not masked by shape-induced anisotropy.} Using a combination of \gls{fdtd} modelling, Mie theory, \gls{qs} and T-matrix analysis, we explore their modal properties and elucidate their unique spectral response.

\section{Results and discussion}
\noindent
The hyperbolic material which comprises the studied nanoparticles is based on an artificial metal-dielectric multilayer to, on one hand, closely relate to the recently studied hyperbolic Au-\ce{SiO2}/\ce{Al2O3}/\ce{TiO2} nanostructures \cite{Maccaferri2019, Isoniemi2020} and, on the other hand, retain the ability to freely tune the birefringence by changing the composition. Thus, we assemble the hyperbolic tensor $\bm{\epsilon}$ by combining a dispersionless dielectric with refractive index $n_d$ and silver \cite{palik_vol1}. 
The nanoresonators, here spherical in shape, are schematically represented in Fig.~\ref{fig:scheme}a with $\epsilon_{xx}= \epsilon_{yy}\equiv \epsilon_\perp$ and $\epsilon_{zz}\equiv\epsilon_\parallel$. Here, the subscripts $\parallel$ and $\perp$ indicate components parallel and perpendicular to the anisotropy axis, respectively, and the tensor elements are given as \cite{Zapata-Rodriguez2013}
\begin{subequations}
\begin{align}
    \epsilon_\parallel&=\frac{\epsilon_m \epsilon_d}{(1-f_m)\epsilon_m+f_m\epsilon_d}, \\
    \epsilon_\perp&=(1-f_m)\epsilon_d+f_m\epsilon_m,
\end{align}
\end{subequations}
where $f_m$ is the filling fraction of metal with permittivity $\epsilon_m$ and $\epsilon_d=n_d^2$.

For a metal fill factor $f_m=0.5$ the exemplary dispersion is plotted in Fig.~\ref{fig:scheme}b, which illustrates the presence of two types of isofrequency surfaces \cite{2013_NatPhoton_7_958_poddubny} with either two (below $\sim$3.4~eV) or one ($\sim$3.4--3.8~eV) permittivity tensor elements being negative. For a larger $f_m$, $\bm{\epsilon}$ tends to that of isotropic silver, while for smaller $f_m$ towards an isotropic dielectric.

Previous research has proven that an effective medium description of a multilayer can be adequate for bulk structures. The former is, naturally, a much more efficient approach to calculate optical properties, however, we need to ensure that this equivalency will remain also for small nanostructures. To that end we first use accurate \gls{fdtd} simulations with mesh sizes down to 5~\AA\/ and layer thicknesses in the range 1--5~nm to confirm their convergence. \corr{For simplicity, we explicitly neglect a number of material-related effects reported in the literature, such as dependence of permittivity of thin layers on their thickness \cite{Laref2014, Stefaniuk2014}, nonlocality \cite{Orlov2011, Sun2015}, quantum Landau damping \cite{Castillo-Lopez2019}, or surface roughness \cite{Andryieuski2014, Kozik2014}. This is motivated by our objective, namely to elucidate the dependence of optical properties of \gls{hns} on the anisotropy of the constituent material, which is assumed to be a given.} 

\begin{figure*}
  \includegraphics[width=0.95\textwidth]{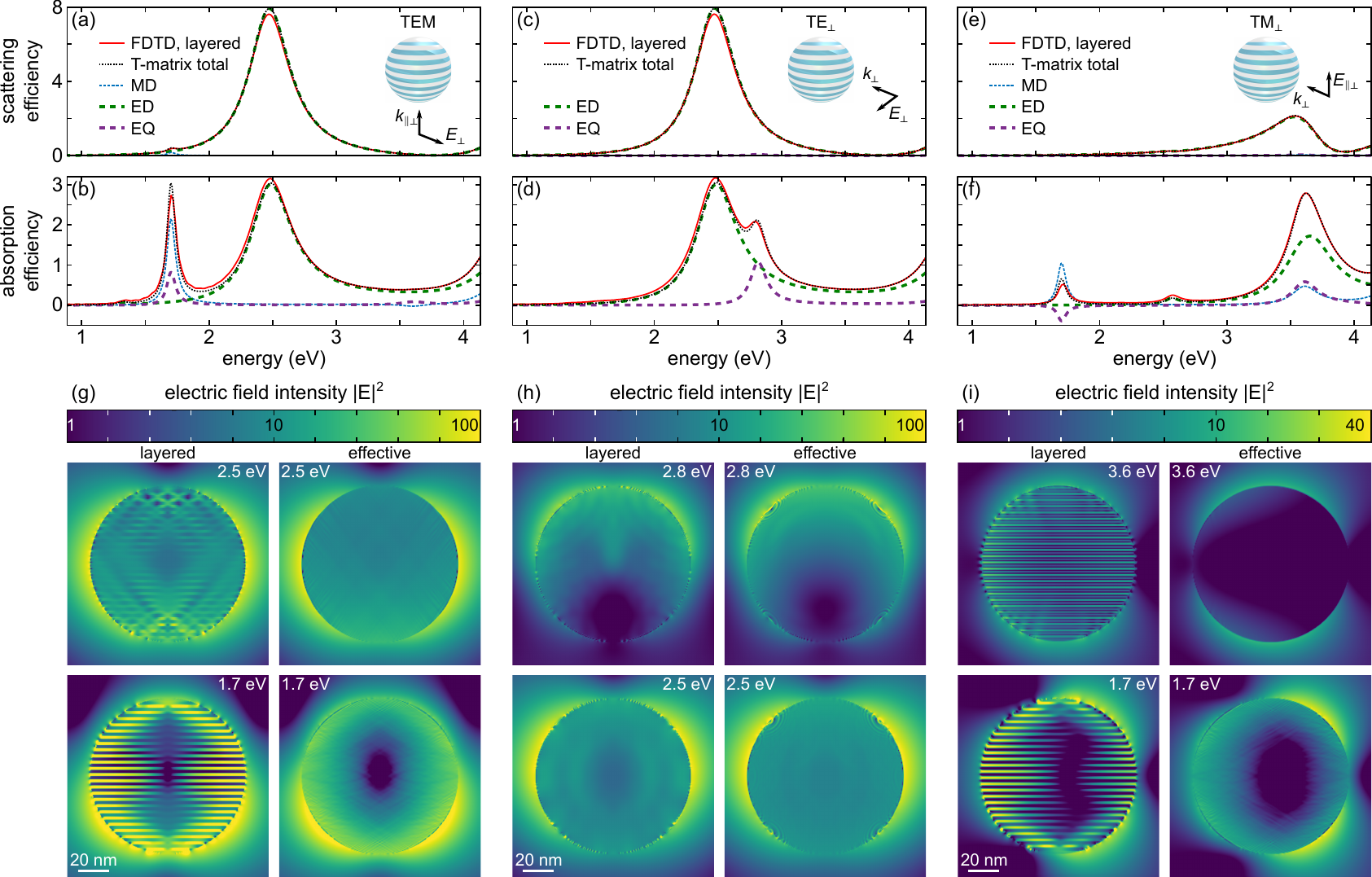}
  \caption{(a-f) Comparison of scattering and absorption of a hyperbolic sphere ($r=50$~nm): the \gls{fdtd} method with explicit Ag-dielectric multilayers ($f_{m}=0.5$) versus the T-matrix method for an effective permittivity. In the three unique incidence/polarization cases the two methods agree quantitatively. Here, three multipoles are active: the \gls{ed}, \gls{md}, and \gls{eq}, and appear at various energies depending on illumination. For example, for $E_\parallel$ a strong \gls{ed} appears at \unit[2.5]{eV}, but is present  above \unit[3.5]{eV} and is mixed with the \gls{md} and \gls{eq}. The \gls{md} is excited only for $k_\perp$, $E_\parallel$ and $k_\parallel$, $E_\perp$ but is, respectively, intermixed with the \gls{eq} in- or out-of-phase. This mixing creates a unique resonance which is very strongly absorptive and has minute scattering.
  (g-i) Comparison of electric field cross sections from \gls{fdtd} and T-matrix calculations at selected resonances further demonstrates good agreement between the two approaches, validating the use of an anisotropic effective permittivity.}
  \label{fig:comparison}
\end{figure*}

The optical spectra (from \gls{fdtd}) for three unique plane wave incident/polarization combinations are then plotted in Fig.~\ref{fig:comparison} for a sphere with radius $r=50$~nm, $f_m=0.5$, and 4~nm layers and compared to T-matrix calculations (open-source code \textsc{smuthi} \cite{Egel2014, Egel2016, smuthi}) for an effective hyperbolic permittivity. The agreement between the scattering and absorption spectra, Fig.~\ref{fig:comparison}a-f, is excellent, and shows only small underestimation of some of the resonance amplitudes. This agreement makes it possible to use, throughout the rest of this work, the more efficient T-matrix method for all considered nanoparticles. Furthermore, the electric field cross sections at selected resonances in Fig.~\ref{fig:comparison}g-i show good agreement as well both inside and outside the \gls{hns}, especially considering the qualitatively different structures.

The benefit of the T-matrix method is that it inherently provides the response of all relevant multipoles and all coupling elements between them. In Fig.~\ref{fig:comparison}a-f we also plot contributions from three active multipoles, namely a strong \gls{ed}, a weaker \gls{eq}, and the lone \gls{md}.

In Fig.~\ref{fig:comparison}a,b the wave vector of the incident field is along the anisotropy axis, making the orientation of the linear polarization irrelevant. In the hyperbolic dispersion range the resonator has a dominant \gls{ed} resonance at $\sim2.5$~eV, which shows up in scattering and absorption, and a second resonance at $\sim1.7$~eV, which consists of a \gls{md} and \gls{eq}. This second resonance is quite unique in that its scattering efficiency is much smaller than its absorption. This is in agreement with the experimental observation made by Maccaferri et al. \cite{Maccaferri2019}.

The second and third illuminations conditions have the incident wave vector perpendicular to the anisotropy axis with the electric field either polarized perpendicular (Fig.~\ref{fig:comparison}c,d) or parallel to the anisotropy axis (Fig.~\ref{fig:comparison}e,f). In the former case only the \gls{ed} and \gls{eq} multipoles are present in a spectral arrangement which is reminiscent of the response of an isotropic plasmonic particle. Also note, that the position of the \gls{ed} matches that of the \gls{ed} in the first case shown in Fig.~\ref{fig:comparison}a,b. In the latter case in Fig.~\ref{fig:comparison}e,f one sees a strong resonance $\sim3.6$~eV which consists of the \gls{ed}, \gls{md}, and \gls{eq}. Additionally, a second absorptive resonance (with practically no scattering) is seen at $\sim1.7$~eV, which is qualitatively similar to the one at the same energy in Fig.~\ref{fig:comparison}a,b. However, an interesting difference is seen in the sign of the contributions of the \gls{md} and \gls{eq} multipoles, which at this particular resonance are excited either in or out of phase. 

\subsection{Origin of dipolar modes of a hyperbolic nanosphere}
\noindent
While the existence of an unusual dipolar response of a type II hyperbolic nanoantenna has been reported earlier \cite{Maccaferri2019} and confirmed above, in this section we elucidate the origin of their nontrivial radiative and non-radiative properties.
For simplicity we limit the analysis to spheres small compared to the wavelength (with $x\equiv kr\ll1$, where $k$ is the wavenumber), and assume vacuum as the surrounding medium. We begin by expanding the internal modes of the particles into plane waves, following Kiselev \cite{Kiselev2002}, to obtain semi-analytical expressions for the T-matrix of the scatterer. This enables us to find the eigenfrequencies of the \gls{ed} and \gls{md} resonances and discuss the scattering albedo of small hyperbolic spheres.

The boundary problem, which relates the incident and surface fields of the T-matrix of an anisotropic sphere, is formulated as $Q$-integrals \cite{Doicu2014}. We briefly recall the formal notation in which each multipole is characterized by three numbers $(\tau,m,l)$. Here $\tau$ denotes the type of the multipole with $\tau=0$ being a magnetic and $\tau=1$ an electric one, $l$ is the order (1 -- dipole, 2 -- quadrupole, etc.) and $m$ is the azimuthal mode number.
The T-matrix for each pair of multipoles is defined as 2x2 blocks given by equation
\begin{equation}
    T_{m_1,l_1,m_2,l_2}=-Q_{m_1,l_1,m_2,l_2}^{1}\left[Q_{m_1,l_1,m_2,l_1}^{3}\right]^{-1}.
\end{equation}

The $(1,1)$ element of the $Q^3$ matrix is an integral of the form
\begin{multline}
    Q^{3,(1,1)}_{m_1,l_1,m_2,l_2}=\frac{ix^2}{\pi} \int \left[ m_r \left(\vec{\hat{r}}\times  \vec{X}^h_{m_1,l_1}\right)\cdot \vec{M}^3_{-m_2,l_2}+\right. \\ 
    \left.\left(\vec{\hat{r}}\times \vec{X}^e_{m_1,l_1}\right) \cdot \vec{N}^3_{-m_2,l_2}\right] dS,
\end{multline}
where $\vec{M}$ and  $\vec{N}$ are the vector spherical wave functions, $\vec{\hat{r}}$ is the unit vector normal to the sphere surface, $\vec{X}^e$ and $\vec{X}^h$ are internal modes of the anisotropic sphere. Changing the first $Q^3$ matrix index leads to exchange of $\vec{M}$ for $\vec{N}$ and vice versa. Changing the second index leads to replacing of $\vec{X}$ internal modes to $\vec{Y}$ internal modes as shown in \cite{Doicu2014}. Expressions for $Q^1$ are similar, but with $(\vec{M}^3,\vec{N}^3)$ replaced by $(\vec{M}^1,\vec{N}^1)$.
The internal modes of the anisotropic sphere are described in terms of plane waves, which are parametrized by angles $(\alpha,\beta)$ of the plane wave wave vector in spherical coordinates. The detailed expressions for those modes are presented in \cite{Doicu2014}.

In general, solutions of both internal modes and $Q$-integrals require numerical integration and matrix inversion techniques. However, a key observation that enables us to reduce the problem complexity is that the number of plane waves required to describe the modes of an anisotropic sphere is limited. It is sufficient to replace the integral over the polar angle by only two polar angles, while the azimuthal integral is carried out analytically and reasonable approximations for small spheres can be obtained,
%We therefore replace the double integral over the polar and azimuthal angles of the plane waves with only two polar angles, while the azimuthal integral is carried analytically.
\begin{equation}
    \int_0^{\pi} f(\beta) \sin{\beta} d\beta \approx \sum_i f(\beta_i) \sin{\beta_i} w_i.
\end{equation}
The chosen angles are $\beta_1=\pi/4.75$, $\beta_2=\pi-\pi/4.75$ and the weight is equal to $w_1=w_2=\pi/2$ as dictated by the Gauss-Legendre quadrature algorithm. Due to symmetry only a single term of the sum has to be evaluated, while the other is identical and the summation becomes trivial.
The integration over the particle surface can be performed analytically once the anisotropic sphere modes are further approximated using Taylor expansion of Bessel functions. After performing the integration, the denominator and numerator of the resulting fractions are expanded into Taylor series. 

\begin{figure}[t]
    \centering
    \includegraphics[width=\columnwidth]{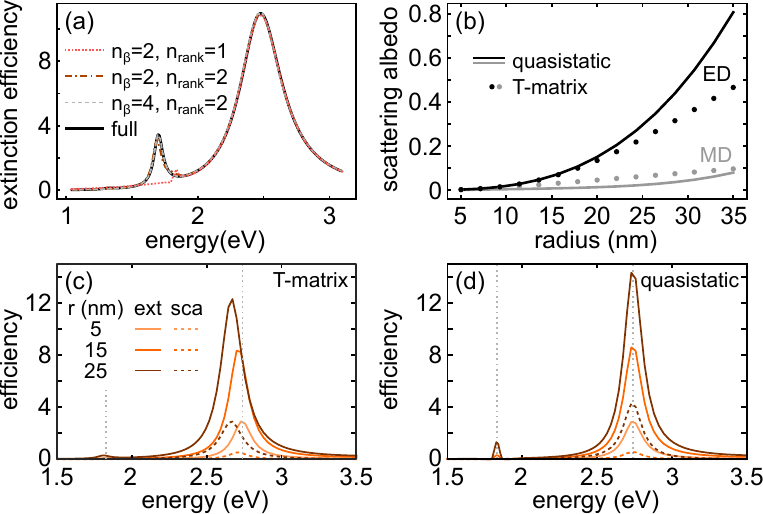}
    \caption{\gls{md} and \gls{ed} dipole comparison between the \gls{qs} and full T-matrix approach. (a) The T-matrix extinction spectrum (black solid line) of a hyperbolic sphere with $r=50$~nm ($k_\parallel$, $E_\perp$) is reproduced correctly for the \gls{ed} and \gls{md} only when accounting for 4 plane waves ($n_\beta=4$) and the first two multipole orders ($n_\mathrm{rank}=2$). (b) The \gls{qs} approximation yields good scattering albedo for the \gls{ed} resonance up to $r\lesssim20$~nm. The \gls{qs} scattering albedo of the \gls{md} is overestimated in a broad size range, however, (c,d) the \gls{qs}-calculated \gls{md} peak  position (and the \gls{ed} one) agrees very well with the T-matrix-derived one.}
    \label{fig:qs-comparison}
\end{figure}

In the limit of small anisotropic nanoparticle size, the electric dipole is not coupled to any other resonance and can be considered in a purely \gls{qs} manner as shown by Bohren and Huffman \cite{Bohren1998}. The resulting expressions for the electric dipole T-matrix components are close analogues of quasi-static Mie theory approximation and the in-plane and out-of-plane permittivities separate:
\begin{subequations}
\begin{align}
    T^{1,1}_{1,1,1,1}&=-\frac{2}{3}i x^3\frac{(-1+\epsilon_\perp)}{2+\epsilon_\perp}\\
    T^{1,1}_{0,1,0,1}&=-\frac{2}{3} i x^3\frac{(-1+\epsilon_\parallel)}{2+\epsilon_\parallel}
\end{align}
\label{eq:eld}
\end{subequations}
Consequently, the dipolar resonance occurs if the following conditions, respectively, are fulfilled
\begin{subequations}
\setlength{\tabcolsep}{0pt}
\noindent\begin{minipage}[b]{0.35\hsize}
\begin{equation}\label{eq.condition1}
    \epsilon_\perp = -2                       \notag
    \addtocounter{equation}{1}
\end{equation}
\end{minipage}
\begin{minipage}[b]{0.10\hsize}
and
\end{minipage}
\begin{minipage}[b]{0.35\hsize}
\begin{equation}\label{eq.condition2}
    \epsilon_\parallel = -2.    \notag
    \addtocounter{equation}{1}
\end{equation}
\end{minipage}\begin{minipage}[b]{0.2\hsize}
\vskip-5pt
\end{minipage}
\hfill(\ref{eq.condition1},\ref{eq.condition2})
\label{eq.both-conditions}
\end{subequations}
\vskip 10pt
\noindent
Equation (\ref{eq.condition1}) corresponds to the incident field with $E_\perp$ for both $k_\parallel$ and $k_\perp$, cf. Fig.~\ref{fig:comparison}a-d and eq. (\ref{eq.condition2}) for $E_\parallel$ and $k_\perp$, cf. Fig.~\ref{fig:comparison}e,f.
As the radius increases, depolarization and radiation increase and the resonance frequency is modified in a similar manner to that of a plasmon resonance in a metallic nanoparticle. 

In contrast to the \gls{ed}, the \gls{md} is more complex. In particles with magnetic permeability $\mu=1$ it can only arise as a consequence of appropriate geometrical structuring, such as large size in dielectric nanoparticles, being essentially a geometrical resonance. Thus, at first glance a \gls{qs} approach is questionable. Indeed, when taken independently, the \gls{md} cannot be accounted for in a \gls{qs} treatment. However, if one follows the results presented in Fig.~\ref{fig:comparison}, namely the simultaneous presence of the \gls{md} and \gls{eq} resonances, and considers the coupled \gls{md}-\gls{eq} peak jointly, the simple treatment is successful. Indeed, Fig.~\ref{fig:qs-comparison}a illustrates that for the magnetic response to show up in the spectral response, at least two multipole orders ($n_\mathrm{rank}=2$) have to be considered. Due to \gls{md}-\gls{eq} coupling, the expression for the magnetic dipole in the \gls{qs} approximation is
\begin{equation}
    T^{0,0}_{1,1,1,1}=0.725i x^5 \frac{ \Psi_1(\epsilon_\perp,\epsilon_\parallel)} {(\epsilon_\perp+1.65 \epsilon_\parallel) \Psi_3(\epsilon_\perp,\epsilon_\parallel)},
    \label{eq:MDT}
\end{equation}
where $\Psi_1(\epsilon_\perp,\epsilon_\parallel)$ \footnotetext[1]{$\Psi_1(\epsilon_\perp,\epsilon_\parallel) = (-1.5+0.25 \epsilon_\perp) \epsilon_\perp^3+ \epsilon_\perp^2 (-4.95+\epsilon_\perp (1.08+\epsilon_\perp)) \epsilon_\parallel+ \epsilon_\perp (-4.1+\epsilon_\perp (1.5 +3.3 \epsilon_\perp)) \epsilon_\parallel^2+\epsilon_\perp (0.68 +2.73 \epsilon_\perp) \epsilon_\parallel^3$}\footnotemark[1] and
$\Psi_3(\epsilon_\perp,\epsilon_\parallel)$ \footnotetext[2]{$\Psi_3(\epsilon_\perp,\epsilon_\parallel)=-16.3 \epsilon_\perp^3+\epsilon_\perp^2 (-49-43.27 \epsilon_\parallel)+ \epsilon_\perp (-81-27 \epsilon_\parallel) \epsilon_\parallel$}\footnotemark[2] are third order polynomial functions of $\epsilon_\perp$ and $\epsilon_\parallel$.

The magnetic resonance observed in Fig. \ref{fig:comparison} and described by eq.~(\ref{eq:MDT}) may occur, when either of the two terms of the denominator is zero. In the present case, the relevant resonance condition is then
\begin{equation}
    \epsilon_\perp+1.65 \epsilon_\parallel = 0.
    \label{eq:mdresonance}
\end{equation}
This equation, qualitatively corresponding to the Fr\"ohlich condition for a small plasmonic sphere [and here cf. eq.~(\ref{eq.both-conditions})], defines a material-type resonance, but has an important difference. Namely, it is only fulfilled provided that the material is hyperbolic. This is markedly different for isotropic nanospheres, in which the \gls{md} is geometrical and is captured analytically only once higher order approximations of the Mie theory are used. In contrast, here it is purely a material resonance, which results from an interplay between dipolar and quadrupolar internal fields also for relatively small particles.

The approximate solutions for the \gls{ed} and \gls{md} can be also used to explain the fact that the electric mode is predominantly radiative, while the magnetic mode is strongly absorbing. The scattering albedo, which is the ratio between scattering and extinction efficiencies, is given by $|T|^2/\mathrm{Re}(T)$. For small spheres the magnetic moment is proportional to $x^5$ according to eq.~\ref{eq:MDT}, while the electric moment is proportional to $x^3$ (Eqs. \ref{eq:eld}). As $x\ll1$, the scattering albedo is damped significantly for the \gls{md} due to small $|T|^2 \propto{x^{10}}$. This is indeed observed when we plot the scattering albedo for hyperbolic spheres in Fig. \ref{fig:qs-comparison}b. 

Further confirmation that this proportionality is the main factor influencing the scattering albedo in the small particle size limit is that in this regime absorption is the dominant extinction channel, even if the material losses are vanishingly small, as shown in Fig. S1. Such behavior is predicted by both QS and full T-matrix calculations. For sufficiently large size particle size scattering becomes the dominant channel when the material losses are low, at which point the scattering albedo is determined by material losses. As a consequence of the proportionality difference (with respect to $x$), the scattering albedo of the \gls{ed} is determined mostly by material losses for much smaller particles than the \gls{md} scattering albedo.

The scattering albedo is also a good measure of the applicability of the \gls{qs} approximation. As the nanoparticle size increases, various radiative effects including depolarization \cite{JOSAB_26_517_moroz} decrease the scattering albedo, ensuring energy conservation of the optical response. When comparing the full T-matrix scattering albedo and the analogous QS method result, we conclude that the applicability range is up to about 20 nm, see Fig.~\ref{fig:qs-comparison}b.

Furthermore, depolarization leads to a shift of the resonance energy. For the smallest nanoparticle size ($r=5$~nm) considered in Figs. \ref{fig:qs-comparison}c-d, the T-matrix and \gls{qs} approaches agree very well in terms of both \gls{md} and \gls{ed} energies and extinction amplitudes proving the validity of \gls{qs} method. However, when depolarization is neglected, the resonance energies are size-independent and the \gls{qs} approximation cannot reproduce the size dependence of resonance energies observed in T-matrix calculations. 

\subsection{T-matrix symmetry for uniaxial spheres}
\noindent
In order to explain the dependence of the optical response on illumination, we utilize the group theory based approach presented in \cite{Schulz1999}, which enables one to find the non-zero entries of the T-matrix.  Then, we study the spherical wave expansion of a plane wave to derive a set of rules that rationalize the observed response for each of the three studied illumination conditions.

A uniaxial sphere belongs to the $D_{\infty h}$ point group having continuous rotation symmetry and a horizontal plane of reflection. Indeed, just based on the symmetry of the nanoparticle it is possible gain fundamental understanding of its mode structure \cite{2020_PRB_102_075103_gladyshev, 2020_OpEx_28_3073_xiong}. Rotation symmetry imposes that the T-matrix is diagonal with respect to $m$
\begin{equation}
    T^{\tau_1,\tau_2}_{m_1,l_1,m_2,l_2}=
    \delta_{m_1,m_2} T^{\tau_1,\tau_2}_{m_1,l_1,m_2,l_2}.
\end{equation}
A given T-matrix element for negative $m$ is the same as for the positive one if the multipole is of the same type. Otherwise, the sign is switched when changing $m$ to $-m$
\begin{equation}
    T^{\tau_1,\tau_2}_{ m,l_1, m,l_2}=
    (-1)^{\tau_1+\tau_2} T^{\tau_1,\tau_2}_{-m,l_1,-m,l_2}.
\end{equation}
For cross-coupling the change of order of interacting multipoles (e.g. \gls{ed}-\gls{mq} to \gls{mq}-\gls{ed}) follows
\begin{equation}
                         T^{\tau_1,\tau_2}_{ m,l_1, m,l_2}=
    (-1)^{\tau_1+\tau_2} T^{\tau_2,\tau_1}_{ m,l_2, m,l_1}=
                         T^{\tau_2,\tau_1}_{-m,l_2,-m,l_1}.
\end{equation}
Finally, the presence of a horizontal plane of reflection imposes 
\begin{equation}
    T^{\tau_1,\tau_2}_{m,l_1,m,l_2}=0,
\end{equation}
if $l_1+l_2 + 2|m|+\tau_1+\tau_2$ is an odd number. 

\begin{figure}[t]
    \centering
    \includegraphics[width=\columnwidth]{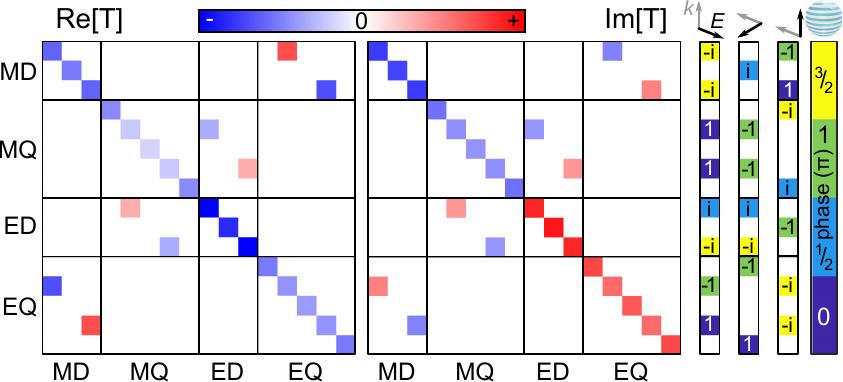}
    \caption{Real and imaginary parts of an exemplary T-matrix of a uniaxial nanosphere (with $D_{\infty h}$ point group) with plotted \gls{md}, \gls{mq}, \gls{ed}, and \gls{eq} and projection of the three unique incident plane waves into these multipoles and their azimuthal modes.}
    \label{fig:t-matrix-example}
\end{figure}

An exemplary T-matrix derived using these rules is presented in Fig.~\ref{fig:t-matrix-example}. We observe that for each block of the diagonal part of the T-matrix are $l+1$ unique elements and that each diagonal element remains unchanged upon changing the sign of azimuthal mode $m$. This makes the following sense: for a dipole there are parallel and perpendicular polarizabilities, for a quadrupole there are parallel-parallel, perpendicular-perpendicular and parallel-perependicular etc.
%The properties resulting from coupling are studied later on. 

Simultaneously, the T-matrix has non-zero off-diagonal elements, which mirror interparticle coupling selection rules \cite{PhysRevB.102.085431}. These can be summarized as follows
\begin{enumerate}
    \item if $\tau_1=\tau_2$: $l_1$ can couple to $l_2$ provided that $l_1+l_2$ is even,
    \item if $\tau_1 \neq \tau_2$: $l_1$ can couple to $l_2$ provided that $l_1+l_2$ is odd.
\end{enumerate}
Stated simply, coupling between electric and magnetic multipoles can happen only if one of their orders is even and the other is odd. For such orders, coupling between multipoles of the same type requires skipping every other order. 

\subsection{Plane wave excitation of uniaxial spheres}
\noindent
To elucidate the spectra of \gls{hns} it is necessary to express the incident field in terms of vector spherical wave functions to show which modes can be excited and with which phase.
Extinction is calculated from a given T-matrix as $\sigma\propto {\rm Re}(\bm{a}^{*}T\bm{a})$, where $\bm{a}$ is a vector of the initial field coefficients in the $(\tau,m,l)$ basis. To facilitate analysis of the mode structure, it is convenient to recast extinction of a given $(\tau,m,l)$ mode into two parts. The first is connected with \textit{pure mode} excitation and the second results from \emph{coupling} between modes with various $\tau$, $m$ and $l$
\begin{equation}
    \sigma_{m,l}^\tau \propto \mathrm{Re} \Big(|{a}_{m,l}^{\tau}|^2 
    \underbrace{ T^{\tau,\tau}_{m,l,m,l}}_{\overset{\text{\scriptsize{pure}}}{\text{\scriptsize{mode}}}} +
    {a_{m,l}^{\tau}}^* \sum_{m',l',\tau'} 
    \underbrace{ T^{\tau,\tau'}_{m,l,m',l'} }_{\text{\scriptsize{coupling}}}
    a_{m',l'}^{\tau'} \Big).
\end{equation}
The above equation underlines the significant influence of coupling on the extinction spectrum. It is determined not only by off-diagonal elements of the T-matrix, but also the incident field coefficients, i.e. $[{a_{m,l}^{\tau}}]^*  a_{m',l'}^{\tau'}$ product. 

Now we focus on plane wave illumination with initial field coefficients $a_{P,m,l}^{\tau}$, where we add the index $P$ to denote the wave's polarization. 
For normal incidence along the optical axis, TE$_\parallel$ with $k_\parallel$ (equivalent to TM$_\parallel$ except for a multiplicative constant $i$; both cases of normally incident light will be referred to as TEM where appropriate) the initial field coefficients are given by \cite{Doicu2014}
\begin{align}
    a^{0}_{\mathrm{TE}_{\parallel},\pm 1, l}&=-i^l \sqrt{2l+1}\label{eq:atepar0}\\
    a^1_{\mathrm{TE}_{\parallel},\pm 1, l}&=-i^{l+1} \sqrt{2l+1}(\mp i)
    \label{eq:atepar1}
\end{align}     
and zero otherwise. When the wave vector of the incident plane wave is perpendicular to the optical axis, there are two cases for the TM$_\perp$ and TE$_\perp$ polarizations. In the first case for TM with $k_\perp$ and $E_\parallel$ we have
% \end{equation}
\begin{align}
  a^0_{\mathrm{TM}_{\perp},m,l}  &= 
  \begin{cases}
    0, & l+|m| \text{ even} \\
    i\frac{(l+|m|)!!\cdot(l-|m|+1)}{(l+1-|m|)!!}\Tilde{c}_{ml}, & l+|m| \text{ odd}
  \end{cases} \label{eq:atmperp0} \\
  a^1_{\mathrm{TM}_{\perp},m,l} &= 
  \begin{cases}
    m\frac{(l+|m|-1)!!}{(l-|m|)!!}\Tilde{c}_{ml}, & l+|m| \text{ even} \\
    0, & l+|m| \text{ odd}
  \end{cases} \label{eq:atmperp1} 
\end{align}
For the TE polarization ($k_\perp$, $E_\perp$)
\begin{align}
  a^0_{\mathrm{TE}_{\perp},m,l} &= 
  \begin{cases}
    -i m \frac{(l+|m|-1)!!}{(l-|m|)!!}\Tilde{c}_{ml}, &  l+|m| \text{ even} \\
    0, & l+|m| \text{ odd}
  \end{cases} \label{eq:ateperp0} \\
  a^1_{\mathrm{TE}_{\perp},m,l}  &= 
  \begin{cases}
    0, & l+|m| \text{ even} \\
    \frac{(l+|m|)!!\cdot(l-|m|+1)}{(l+1-|m|)!!}\Tilde{c}_{ml}, & l+|m| \text{ odd}
  \end{cases} \label{eq:ateperp1}
\end{align}
The above used $\Tilde{c}_{ml}$ constant is derived and defined in the \gls{si}.
Using the above coefficients we can discuss how the excited modes depend on the incident field polarization and wave vector direction.

\begin{figure*}
  \includegraphics[width=18cm]{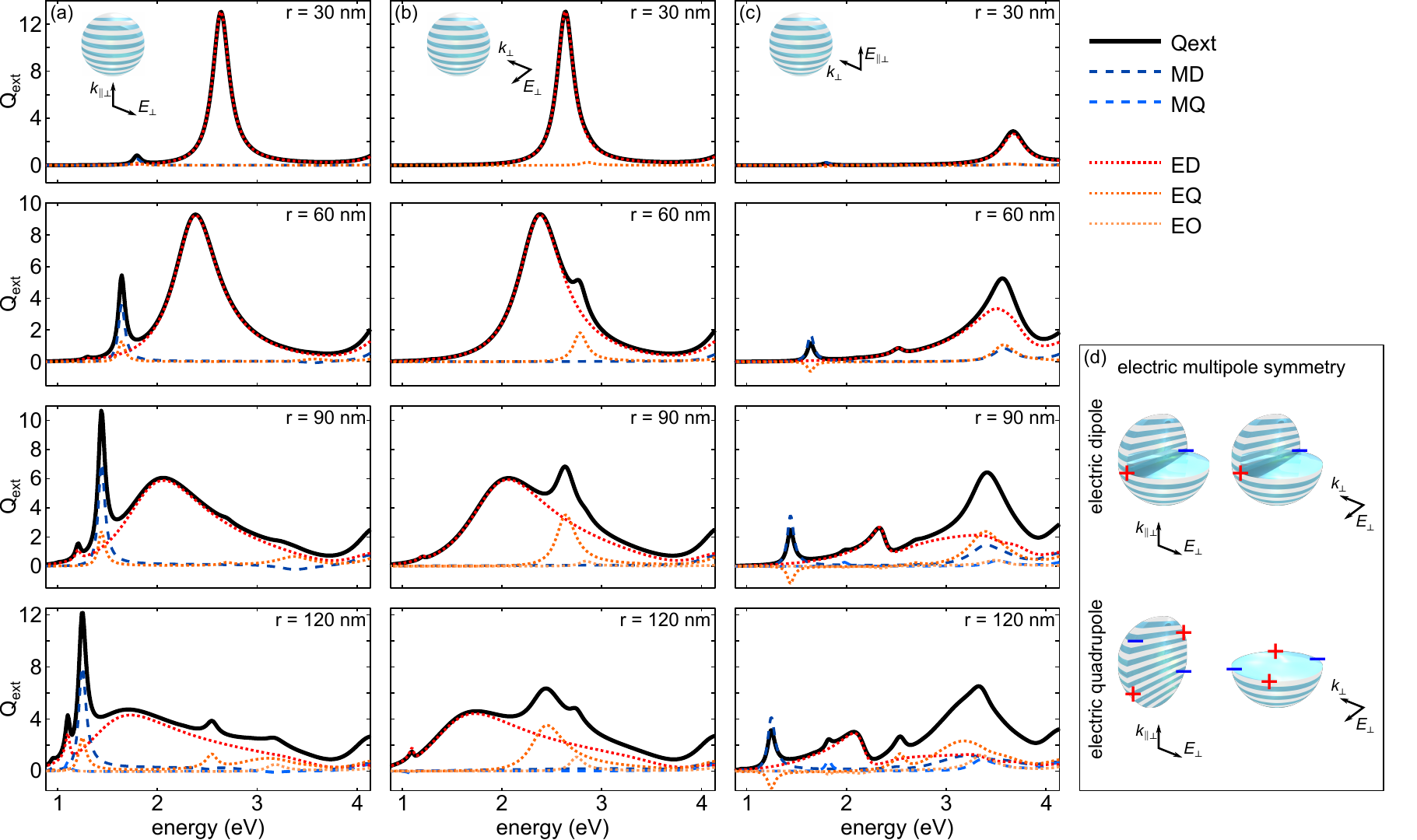}
  \caption{Spectra of total extinction efficiency (solid black line) and their decomposition into five moments \gls{md}, \gls{mq}, \gls{ed}, \gls{eq}, and \gls{eo} vs illumination orientation and hyperbolic ($f_{m}=0.5$) sphere radius. (a) For $k_\parallel$ (TEM polarization) for small $r$ the \gls{ed} is dominant, but with increasing $r$ a very sharp \gls{md} appears which is coupled with the \gls{eq}. However, higher electric multipoles remain weak even for $r=120$~nm, while an additional \gls{ed} appears below the \gls{md}. (b) For $k_\perp$ and $E_\perp$ (TE$_\perp$ polarization) the spectra evolve with $r$ as those of a pure metallic nanosphere with first a dominant \gls{ed}, followed by a stronger \gls{eq} for $r=90$~nm, and then a \gls{eo}. The \gls{md}, which is present in case (a) is absent here. (c) For $k_\perp$ and $E_\parallel$ (TM$_\perp$ polarization)  the spectrum is of smaller amplitude and qualitatively different. The \gls{md} peak is at the same spectral location as for case (a), but couples out-of-phase to the \gls{eq}. The electric multipoles below \unit[3]{eV} grow gradually with $r$. Interestingly, the preak above \unit[3]{eV} is composed of a mix of all multipoles with higher orders appearing with increasing $r$. (d) Induced electric multipoles for TEM and TM$_\perp$ polarizations exhibit different symmetry in relation to the anisotropy axis, illustrating the same response of the \gls{ed} mode and different response of the \gls{eq} mode.}
  \label{fig:var-radius}
\end{figure*}

Regardless of polarization and wave vector, both electric and magnetic multipoles are always excited and all orders enter the expansion, while the azimuthal modes are governed by the zeros of the Legendre polynomials and their derivatives as shown in the SI. 
For normal propagation, $k_\parallel$, only modes with $|m|=1$ can be excited. Alternatively, for  $k_\perp$ the selection rules can be deduced as follows. First, assuming TE$_\perp$ polarization:
\begin{enumerate}
    \item if $l$ is odd, even azimuthal modes are excited for magnetic modes and odd ones for electric modes;
    \item if $l$ is even, odd azimuthal modes are excited for magnetic modes and even ones for electric modes.
\end{enumerate}
For TM$_\perp$ polarization the opposite parity holds.
The phase of the incident field for various polarizations is described by the relations
\begin{equation}
    \bm{a}_\mathrm{TE}^{0} = -i \bm{a}_\mathrm{TM}^{1} \quad \text{and} \quad \bm{a}_\mathrm{TE}^{1} = -i \bm{a}_\mathrm{TM}^{0}.
\end{equation}
Additionally, for normal incidence $|\bm{a}_{\mathrm{TE}_\parallel}|=|\bm{a}_{\mathrm{TM}_\parallel}|$, 
because of the relation $a_{P_\parallel,m,l}^{0} = \text{sgn}(m) a_{P_\parallel,m,l}^{1}$. The phase for the three unique polarizations is shown in Fig.~\ref{fig:t-matrix-example}. 
Moreover, only an incident field with $k_\perp$ (TE$_\perp$/TM$_\perp$) with an azimuthal mode $m=\pm 1$ can be expressed in relation to normal incidence ($k_\parallel$) as
\begin{equation}
    a_{P_\perp,\pm1,l}^{\tau} = (-1)^{l-1} a_{P_\parallel,\pm1,l}^{\tau}.
    \label{eq:a2}
\end{equation}
 In practice, one can express $\mathrm{TE}_\perp$-polarized electric dipole and magnetic quadrupole as well as $\mathrm{TM}_\perp$-polarized magnetic dipole and electric quadrupole incident field coefficients in terms of normal incidence incident field coefficients.

 %Higher orders exhibit similar behavior only if terms with $|m|>1$ are negligible. 

We shall now discuss the extinction spectra of hyperbolic nanospheres using the above relations while simultaneously investigating how they evolve with radius. We plot extinction spectra for increasing $r$ from 30 to \unit[120]{nm} in Fig.~\ref{fig:var-radius} and in the \gls{si} for the three unique incidence/polarization cases.

For polarization $E_{\perp}$ and irrespective of $\bm{k}$ direction, a dominant \gls{ed} is always present and of the same amplitude at a given $r$, cf. Fig.~\ref{fig:var-radius}ab, as would be expected for an isotropic plasmonic nanosphere. This can be rationalized by the fact that for $E_{\perp}$ polarization only $m=\pm1$ modes are excited [eq. (\ref{eq:atepar1}), eq. (\ref{eq:ateperp1})] and \gls{ed} is negligibly coupled to other resonances. However, the higher electric multipoles are only visible in the plots in Fig.~\ref{fig:var-radius}b with $k_{\perp}$, while being much weaker for $k_\parallel$. Focusing on the \gls{eq} around 2.5~eV as an example, its amplitude for $\mathrm{TE}_\perp$ polarization at $r=60$~nm is larger than for normal incidence for $r=120$~nm. In fact, the spectral response of the \gls{hns} under $\mathrm{TE}_\perp$ illumination shows a plasmonic-like response, while not so for TEM illumination. This qualitative difference between the two cases can be understood as follows. 

The electric dipole consists of positive and negative charges induced at opposite ends of the nanosphere, as schematically illustrated in Fig.~\ref{fig:var-radius}d. If one neglects retardation and simplifies the \gls{ed} to a positive and negative induced point charge, then the induced \gls{ed} due to symmetry for TEM and $\mathrm{TM}_\perp$ polarizations in relation to the anisotropy is identical.  
From an energy point of view, the \emph{self-energy} of the two \glspl{ed} is the same. 
However, for the quadrupoles the circumstances are different. A physical \gls{eq} consists of four point charges which are located in the plane of incidence as marked schematically in  Fig.~\ref{fig:var-radius}d. For TEM incidence the \gls{eq} is arranged along the anisotropy axis, while for $\mathrm{TE}_\perp$ it is perpendicular to it. Hence, due to the different symmetries of these two cases, the origin of the qualitative difference between the two \gls{eq} is clear.

The other significant difference between the TEM and $\mathrm{TE}_\perp$ cases is the presence of a strong \gls{md} for $k_\parallel$ (Fig.~\ref{fig:var-radius}a), which is weakly radiative for small $r$ as discussed above. Neither the \gls{md} nor other magnetic modes are present in Fig.~\ref{fig:var-radius}b ($\mathrm{TE}_\perp$), when the spectra of a \gls{hns} are reminiscent of an isotropic plasmonic nanosphere for all radii (and $\lesssim$\unit[3.5]{eV}). Magnetic modes are, however, present in Fig.~\ref{fig:var-radius}c ($\mathrm{TM}_\perp$), specifically the magnetic one is found at the same spectral location as in Fig.~\ref{fig:var-radius}a.  However, its amplitude is much weaker due to destructive interference from the coupled \gls{eq}. Conversely, in Fig.~\ref{fig:var-radius}a the \gls{md} and \gls{eq} interfere constructively (cf. Fig.~S3).

The origin of this different interaction between the \gls{md} and \gls{eq} modes for TEM ($k_\parallel$) and $\mathrm{TM}_\perp$ polarizations ($k_\perp$, $E_\parallel$) is deduced from the T-matrix and the incident field, which are depicted in Fig.~\ref{fig:t-matrix-example}. In both cases, the incident field excites the same azimuthal $m=\pm1$ modes of both the \gls{md} and \gls{eq}, which would result in the same extinction in the absence of \gls{md}-\gls{eq} coupling. However, \gls{md}-\gls{eq} coupling contributes to extinction with different sign for $k_\parallel$ and $k_\perp$ TE incidence as predicted by eq.~(\ref{eq:a2}). This same eq.~(\ref{eq:a2}) also explains the switch between constructive/destructive \gls{ed}-\gls{mq} coupling illustrated in Fig.~S3, clearly visible for $r\gtrsim80$~nm. Detailed derivation of the sign of the electric or magnetic quadrupoles' contribution is presented in section S2 of the \gls{si}. 
Hence, based on the spectra in Figs.~\ref{fig:var-radius} and S3, it is clear, that the multipolar properties of even the most simple of hyperbolic nanoparticles, i.e. a sphere, are quite complex. In particular, due to cross coupling of electric and magnetic modes, higher order modes can appear at energies below lower order modes of the same type.

\corr{The negative contribution of the \gls{eq} to extinction and absorption deserves a qualitative discussion. For any system composed of coupled elements its total extinction is positive. However, due to coupling (multiple scattering) within the system some resonators may receive more energy than is directly provided to them by the source. If the coupling is large enough and specifically out-of-phase with the incident field, this particles' extinction may turn negative, implying that it effectively returns more energy to the electromagnetic field than it receives directly from the source \cite{JPCC_116_20522_tja}. 
While this phenomenon elucidates negative extinction in an element, it does not apply to absorption in individual (even coupled) resonators, in which it has to be positive, in particular in a \gls{hns} (cf. Fig.~S3). However, when the optical cross sections are decomposed into multipolar components, scattering of every multipole is always positive \cite{Doicu2014}. On the other hand, extinction of individual multipoles does not have to be larger than scattering of the same multipoles or even be positive, since they are expressed by the expansion coefficients of both the incident and scattered fields \cite{Doicu2014}.
Thus, in a resonator without spherical symmetry coupling between multipoles may beget very efficient energy transfer between them, what from the outside may look like  particular mode is ``generating'' energy. This phenomenon, like negative extinction \cite{JPCC_116_20522_tja}, is connected with out-of-phase coupling between the interacting multipoles, as derived in section S2 of the \gls{si}.}

% Moved original text by MB and TA to stash.tex at position #1.

\begin{figure}
  \includegraphics[width=8.15cm]{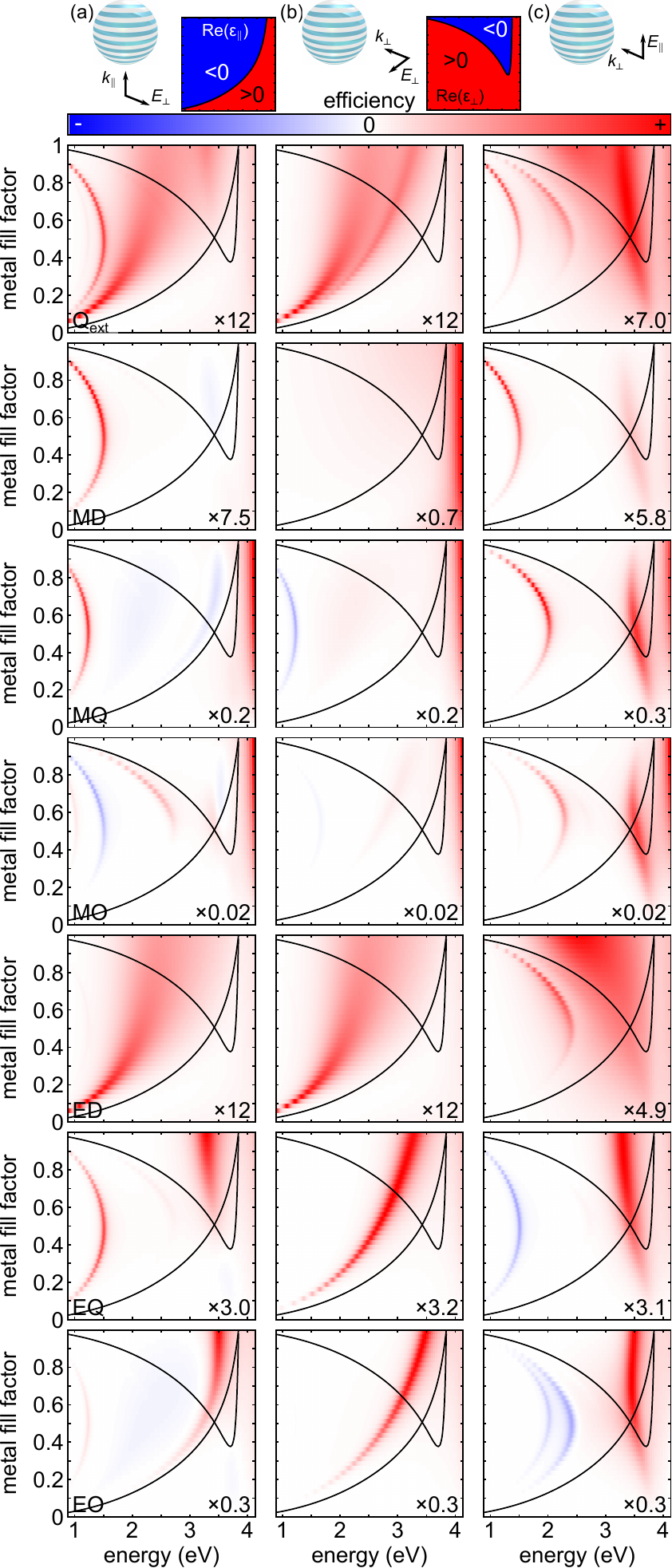}
  \caption{Optical properties of hyperbolic nanospheres, $r=80$~nm, for the three unique illumination conditions vs energy and metal filling fraction: extinction efficiency, $Q_{ext}$, its decomposition into the first six moments: \gls{md}, \gls{mq}, \gls{mo}, \gls{ed}, \gls{eq}, \gls{eo}. Black lines mark regions with positive/negative permittivity tensor elements. Note the monotonic behavior of the mainly uncoupled electric modes with $f_m$, while the magnetic resonances in the hyperbolic dispersion range in all cases coupled to other modes. Furthermore, often the coupled modes of high order occur below the lower order modes of the same type, e.g. \gls{eq} below \gls{ed} for TEM polarization.}
  \label{fig:var-ff80}
\end{figure}

\subsection{Dependence on material properties\label{sec-mat-properties}}
\noindent
Up to this point the hyperbolic material was composed of equal amounts of metal and dielectric with $f_m=0.5$. However, as shown in Fig.~\ref{fig:scheme}cd, varying $f_m$ allows for significant tunability from a uniaxial dielectric through a type I or II hyperbolic material to a uniaxial metal. In Fig.~\ref{fig:var-ff80} we present how the extinction spectrum and its multipole decomposition evolve with $f_m$ for a \gls{hns} with $r=80$~nm.

It is clearly seen, that overall the dominant response is of the electric type, with the \gls{ed} and \gls{md} exhibiting a strong response, especially for $\mathrm{TE}_\perp$ polarization in Fig.~\ref{fig:var-ff80}b, which has the typical response of a plasmonic sphere. This can be confirmed by plotting the spectra vs $\sqrt{f_m}$, which show an almost linear dependence consistent with the localized surface plasmon frequency, as well as the \gls{eq} and \gls{eo}, being proportional to the carrier concentration.
Also note, that while initially absent for small $f_m$, the \gls{eq} and \gls{eo} resonances appear for TEM and $\mathrm{TM}_\perp$ polarizations when the material becomes a uniaxial metal.

\begin{figure}
  \includegraphics[width=8.5cm]{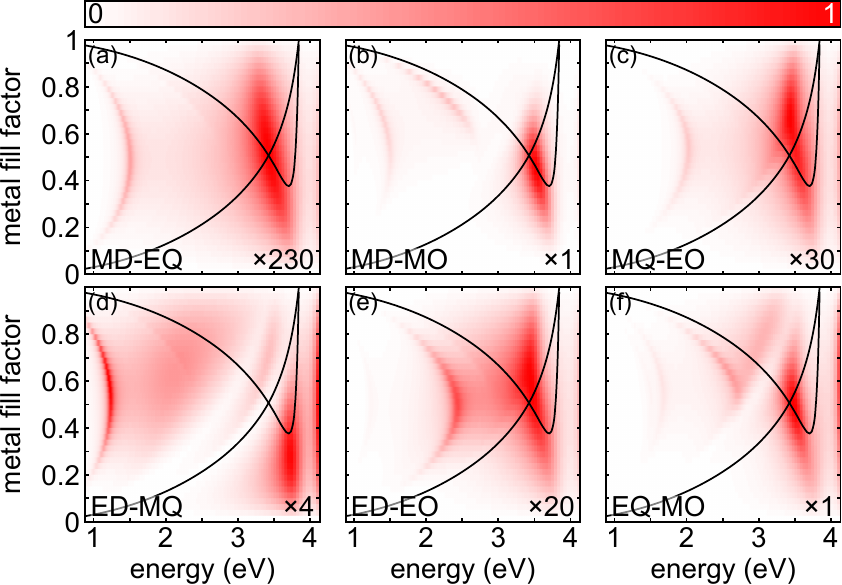}
  \caption{T-matrix derived coupling between multipoles in an anisotropic nanosphere with $r=80$~nm and $n_d=1.5$ whose extinction is plotted in Fig.~\ref{fig:var-ff80}. The actual coupling between multipoles in Fig.~\ref{fig:var-ff80} depends on the incident field direction and polarization. }
  \label{fig:var-ff80-couple}
\end{figure}

The \gls{md} response is, however, quite strong and clearly seen in for  Fig.~\ref{fig:var-ff80}ac. Furthermore, in contrast to the monotonic dependence of the main electric resonances, the \gls{md} and higher magnetic order modes show a peculiar crescent-like shape. Indeed, the electric modes directly coupled to these crescent-shaped magnetic resonances show identical behavior, while simultaneously exhibiting constructive/destructive interference depending on the polarization of incident light.  The magnitudes of these coupling relations are plotted in Fig.~\ref{fig:var-ff80-couple}.

A peculiar feature of the coupled electric and magnetic modes is the symmetric crescent-like profile with respect to $f_m=0.5$. 
This dependence of the modes on $f_m$ can be rationalized using our analytical model using the electric and magnetic dipoles as examples, while for higher orders the behavior will be qualitatively similar. 
To facilitate the analytical analysis we substitute the experimental dispersion of silver with a Drude permittivity fitted to that of silver
$\epsilon_m(\omega)=\epsilon_\infty - \omega_p^2(\omega^2+i \gamma \omega)^{-1}$ with parameters $\epsilon_\infty=4.18$, $\omega_p=8.76$~eV, $\gamma=91$~meV.
The material data is then substituted into eq.~(\ref{eq:mdresonance}). Assuming negligible influence of material losses on the resonance condition, we solve for the zeroing of the real part of eq.~(\ref{eq:mdresonance}) to find the resonance condition of the magnetic dipole 
%\begin{multline}
%    \omega^\mathrm{MD}_{res}=\omega_p \sqrt{f_m(f_m-1)}\Big(\epsilon_\infty f_m(f_m-1) - \\ 
%0.025\epsilon_d\Big[ 53 +40 f_m(f_m-1) +\\
%    7.28 \sqrt{53. +80 f_m(f_m-1) }\Big]\Big)^{-1/2}.
%\end{multline}
\begin{equation}
    \omega^\mathrm{MD}_{res}\approx\frac{\omega_p \sqrt{F_m}}{\sqrt{\epsilon_\infty F_m - \epsilon_d(1.3 + F_m + 1.6 \sqrt{0.7 + F_m })}},
%    \omega^\mathrm{MD}_{res}\approx\frac{\omega_p \sqrt{F_m}}{\sqrt{\epsilon_\infty F_m - \epsilon_d(1.325 + F_m + 1.63 \sqrt{0.66 + F_m })}},
\end{equation}
where $F_m=f_m(1-f_m)$ is a symmetric function with respect to $f_m=0.5$. The above result is thus consistent with the symmetric, crescent-shaped of the \gls{md} resonance confirms previous result as well as offers further proof of the validity of \gls{qs} approximation. 

An analogous approach yields the resonant frequencies of the electric dipole for perpendicular
\begin{equation}
   \omega_{res} = \frac{\omega_p \sqrt{f_m}}{\sqrt{2-\epsilon_d (-1+f_m)+\epsilon_\infty f_m}}
\end{equation}
and parallel
\begin{equation}
   \omega_{res} = \frac{\omega_p \sqrt{2 + \epsilon_d  -2f_m }}{\sqrt{\epsilon_\infty (2 + \epsilon_d  -2f_m )+2\epsilon_d f_m}}
\end{equation}
polarizations. 
Both approach the well-established result for a quasi-static sphere $\omega_{res} = \omega_p/\sqrt{2+\epsilon_\infty}$ in the limit of purely metallic particles and confirm the earlier observation on the quasi-linear dependence of the resonance positions of the \gls{ed} on $\sqrt{f_m}$. 
Also, for both the \gls{ed} and \gls{md} modes, the eigenfrequencies are proportional to $\omega_p$ of the metal, which indicates that changing the free charge concentration of the conducting material is a direct method of tuning the properties of hyperbolic nanoparticles. 

\corr{\subsection{General considerations on hyperbolic nanoparticles and their permittivities}}
\noindent
\corr{Having obtained the main goal of elucidating the optical properties of \gls{hns}, we briefly discuss a few aspects of hyperbolic materials which could be the basis for realizing hyperbolic nanoresonators.
The degree of anisotropy in known natural materials was typically small until the discovery of \gls{vdw} materials, while natural hyperbolic material were not common. However, with the rise of extensive research on \gls{vdw} materials hyperbolic materials can easily be found in the literature. Examples include \gls{hbn} with two restrahlen bands \cite{2014_NatCommun_5_5221_caldwel} or materials with Drude-like dispersion for the in-plane components, such as considered here, \ce{TaS2} or \ce{TaSe2} \cite{Beal1975, Yan-Bin2007}. While the permittivity of a \gls{vdw} material depends on the number of layers, once the thickness exceeds a few tens of layers the bulk properties are established and the below-discussed challenges are alleviated. However, to gain independence from chemistry and arbitrarily (though within physical limitations) shape the hyperbolicity, the method of choice is to use structured materials in the form of metal-dielectric metamaterials formed into multilayers or wire-media \cite{2013_NatPhoton_7_958_poddubny}. Sizes of these layers (or wires) have to be small enough so that an effective medium approximation will hold, however, using thin layers can bring about certain changes. These encompass issues related to fabrication, fundamental physical effects, or both.}

\corr{One important aspect is that fabrication of thin, continuous layers is challenging, especially for metals. One of the best plasmonic metals, Ag, is known for its island growth and wetting layers are needed to obtain smooth, continuous metal layers, with germanium being a prime choice \cite{Stefaniuk2014}. This enables deposition of sub-10 nm layers with good qualitites and, if needed, Ge can be used as the dielectric multilayer pair to Ag to form low-loss hyperbolic nanostructures with resonances below $\sim1$~eV. As we show in Fig.~S4, already 10~nm layers are adequte to obtain well-formed both \gls{ed} and \gls{md}-\gls{eq} resonances. Directly connected with deposition of thin layers is a question of the achievable surface roughness \cite{Andryieuski2014, Kozik2014}, which can quickly destroy the hyperbolic-material-dependent \gls{md}-\gls{eq} resonance. Already a \gls{rms} of 0.4~nm for 4~nm Ag layers is enough to smear the \gls{md}-\gls{eq} (Fig.~S5). However, by controlling the temperature during metal evaporation an \gls{rms} of 0.2~nm for a 10 nm Ag film is obtainable, and is in fact limited only by the surface roughness of the substrate \cite{Stefaniuk2014a}. Such small \gls{rms} values are enough to retain all the important spectral features of our hyperbolic nanoresonators. Moreover, for thicker layers the impact of \gls{rms} is lower and for 10~nm Ag layers with an \gls{rms} of 0.6~nm the \gls{md}-\gls{eq} peak is easily seen (Fig.~S6). This means that widely used \gls{pvd} and \gls{ald} methods are viable fabrication paths. \Gls{pvd} can be used in a standard lithography process to deposit the layered nanoparticle through a mask \cite{Maccaferri2019}, while both \gls{pvd} and \gls{ald} enable deposition of wafer-scale multilayers for subsequent sacrificial etching of nanostructures \cite{Wang2003, Verre2017}. Such a lithographic approach will yield structures with flat top and bottom surfaces with the simplest resonator being a disk. However, the optical spectra of a hyperbolic nanosphere and nanodisk are qualitatively similar (Fig.~S7). These spectra prove that the general behavior of hyperbolic resonators is well described by eqs. (\ref{eq.both-conditions}) and (\ref{eq:mdresonance}), but the shape dependence is also a key parameter in determining the actual response. Finally, it is necessary to account for the fact that the permittivities of many materials deposited as thin layers differ from their bulk values \cite{Lehmuskero2007, Laref2014, Stefaniuk2014}. }

\corr{Nonlocality, that is the dependence of material properties on the wave vector, can modify the effective medium permittivity depending on the angle of incidence \cite{Sun2015}. In the region of epsilon near zero, additional light waves \cite{Orlov2011} or complex eigenmodes \cite{Orlov2013} can be observed, which are neglected by a local effective medium theory. However, the optical modes reported herein are observed far from the epsilon near zero range and are not expected to play a role in our case. Other potentially relevant effects are tied to the characteristic sizes of metal grains or layer thicknesses, when the movement of charge carriers is inhibited or altered.  This phenomenon is observed for small dimensions of metals which modify the movement of free electrons such as size quantization or surface screening. These lead to spectral shifts and broadening of the surface plasmon \cite{NJP_15_083044_carmina, OpEx_22_24994_carmina} or quantum Landau damping in thin metal layers \cite{Castillo-Lopez2019}.}

\corr{Despite the many causes of why engineered metal-dielectric multilayers may behave differently than modelled using a local approach, the main observations on the origin of the various optical resonances will hold. They may, of course, occur at shifted frequencies due to different hyperbolic permittivity after accounting for the various above mentioned effects. However, such changes are predictable beforehand and many of them can be circumvented by using \gls{vdw} materials which are anisotropic or even hyperbolic and can easily be tuned after fabrication by electrostatic gating \cite{Munkhbat2020}.}

\section{Conclusions and outlook}
\noindent
In this work we have presented a detailed analytical and numerical study of the optical properties of hyperbolic nanospherical antennas using an artificial silver-dielectric effective multilayer as the exemplary material. It is clear that hyperbolic dispersion enables a rich modal structure which is strongly dependent on the polarization and direction of incident light. For $\mathrm{TE}_\perp$ illumination the response mirrors that of a plasmonic nanosphere with a scaled plasma frequency that is determined by the density of charge carriers, exhibiting the full spectrum of multiple electric multipoles up to the number determined by the nanosphere diameter. However, for TEM incidence only the electric dipole remains of plasmonic behavior, while below the \gls{ed} a very strongly absorptive magnetic dipole is present. The \gls{md} is also present for the $\mathrm{TM}_\perp$ polarization, although its optical cross section is much weaker than for the TEM case.

These modal properties are a consequence of the interplay of the T-matrix derived coupling conditions between various multipole orders, whose efficiencies are determined by the hyperbolic dispersion. The unique coupling, which is absent in isotropic nanospheres, begets the appearance of an atypical modal order. For example, very sharp electric quadrupoles occur at lower energies than the first electric dipole resonance. Similar relations are present for the magnetic response.

Furthermore, by employing a quasistatic analysis of the T-matrix of hyperbolic nanospheres we are able to elucidate the origin of the electric and magnetic dipolar modes. With this approach we derive \corr{material-dependent} resonance conditions for the \gls{ed} and \gls{md} in \corr{eqs. (\ref{eq.both-conditions}) and (\ref{eq:mdresonance}), respectively}. Specifically, we prove that the unique \gls{md} mode present in the hyperbolic nanospheres is a material resonance determined by the ordinary and extraordinary permittivities and requires the sign of these two values to be opposite. \corr{It is expected, especially for the plasmon-like \gls{ed}, that the conditions expressed by eqs. (\ref{eq.both-conditions}) and (\ref{eq:mdresonance}) will need to be amended to account for nonspherical resonators by using a shape factor $L$.}
The \gls{qs} approximation is also crucial in explaining recent observations \cite{Maccaferri2019} of why the magnetic dipole is very strongly absorptive, while in contrast the electric one radiates much more efficiently. 
Our analysis shows that the origin of this unusual behavior of the \gls{md} stems from complex coupling between electric and magnetic multipoles, which leads to very strongly scattering or absorbing modes depending on antenna size and dissipative losses.

Finally, we show how the optical response of hyperbolic nanoparticles can be tuned by varying the charge carrier concentration which sets the magnitude of the metallic permittivity tensor. One interesting example is the ability to tune the spectral separation between the electric and magnetic dipoles by varying the plasma frequency. \corrmb{The tunability of hyperbolic nanostructures' spectral response has been studied recently, e.g., in terms of anomalous scattering leading to electromagnetic cloaking \cite{Zapata-Rodriguez2016}. Hyperbolic nanocavities are of interest in the field of strong light-matter interaction; they have been demonstrated to enhance far-field 
radiation and shorten lifetime of coupled quantum emitters \cite{Indukuri2019}. Hyperbolic dispersion was also shown to increase sensitivity of refractometric sensors due to excitation of high-\textit{k} modes \cite{Sreekanth2016}, making such materials viable candidates for refractometric sensing purposes.}
Thus, we are convinced that our study offers critical insight into the electromagnetic properties of hyperbolic nanoparticles, which seem to be extremely promising candidates for novel devices enabling efficient light-matter interaction.

\acknowledgements
\noindent
We acknowledge support by the Polish National Science Center via the projects 2017/25/B/ST3/00744 (K.M.C., D.\'S., T.J.A.) and 2019/34/E/ST3/00359 (M.B.).
% and 2019/35/B/ST5/02477. 
The computations were enabled by resources provided by the Interdisciplinary Center for Mathematical and Computational Modelling  via the project~\#G55-6.

%\end{acknowledgement}

%\begin{suppinfo}
%This will usually read something like: ``Experimental procedures and characterization data for all new compounds. The class will automatically add a sentence pointing to the information on-line:
%\end{suppinfo}

%\bibliography{bib-database}

\end{document}